\documentclass[preprint,aps,prl,showpacs]{revtex4-1}
\usepackage{graphicx}

\usepackage[pdftex,pdfpagelabels]{hyperref}
\usepackage[alsoload={hep,accepted}]{siunitx}

\graphicspath{{./figures/}} 

\begin{document}

\title{A low power photoemission source for electrons on liquid helium}

\author{S. Shankar}
\author{G. Sabouret}
\author{S. A. Lyon}

\affiliation{Dept. of Electrical Engineering, Princeton University, Princeton, NJ 08544, USA}

\begin{abstract}
Electrons on the surface of liquid helium are a widely studied system that may also provide a promising method to implement a quantum computer. One experimental challenge in these studies is to generate electrons on the helium surface in a reliable manner without heating the cryo-system. An electron source relying on photoemission from a zinc film has been previously described using a high power continuous light source that heated the low temperature system. This work has been reproduced more compactly by using a low power pulsed lamp that avoids any heating. About \num{5e3} electrons are collected on \SI{1}{\cms} of helium surface for every pulse of light. A time-resolved experiment suggests that electrons are either emitted over or tunnel through the \SI{1}{\eV} barrier formed by the thin superfluid helium film on the zinc surface. No evidence of trapping or bubble formation is seen.
\end{abstract}

\pacs{07.77.-n; 73.20.-r; 03.67.Lx}

\maketitle

\section{Introduction}
\label{sec:introduction}

Electrons floating on the surface of liquid helium have been extensively used in experiments in the past to study the physics of two-dimensional electron systems~\cite{Ando1982,Andrei1997}. Recently, the charge~\cite{Platzman1999,Dykman2003,Dahm2002,Papageorgiou2005} or spin~\cite{Lyon2006,Sabouret2008} state of electrons floating on superfluid helium has been proposed as an experimentally realizable qubit for quantum information processing. All such experiments require a source to deposit electrons on the liquid helium surface. Such a source would ideally satisfy the following requirements: it must generate the required number of electrons on the helium surface in a reliable and controlled manner; electrons generated should have a low energy so that they do not punch through the energy barrier at the liquid helium surface and form bubbles in the helium; finally, the electron generation process should involve low power inputs into the cryogenic system, so as not to heat it. In this paper, we describe an electron source that satisfies these requirements.

In order to deposit electrons on the helium surface, two schemes have been commonly used -- thermionic emission from a filament~\cite{Rees2010,Papageorgiou2005,Paalanen1985,Williams1971}, and electrons produced by a gas discharge~\cite{Sommer1964,Sommer1971,Klier2000}. Both schemes have significant limitations. Thermionic emission suffers from the disadvantage of heating up the low temperature system. Additionally the emission is difficult to control, often producing too many electrons. Similarly, striking an arc to cause a discharge in the helium gas produces high energy electrons which need to be slowed using additional guard electrodes. The discharge also requires the gas to be at high pressure and therefore a high temperature (\SIrange[repeatunits=false,trapambigrange=false]{1}{2}{\kelvin}). Thus, electron generation requires the system to be heated up from \SI{300}{\milli\kelvin} or below where most of these experiments are performed.

An electron source has been previously built~\cite{Wilen1985} by Wilen and Giannetta using photoemission from a zinc film by ultraviolet (UV) light from a high power continuous xenon arc lamp. Though found to be reliable, the high power lamp caused some heating of the low temperature system. A further question was also not answered. Since the experimental temperature is below the superfluid transition, the zinc source is covered by a thin film of helium whose role in the photoemission process was unclear. Wilen and Giannetta~\cite{Wilen1985} conjectured two possibilities. First, the high power input by the lamp might have been locally boiling off the helium film, allowing electrons to be emitted directly from the zinc surface. Secondly, the electrons emitted from the zinc might have been trapped in the helium film as electron bubbles, and then released.

In this work, we have implemented a similar reliable and easily controlled photoemission source for electrons on helium, using a low power pulsed xenon lamp to eliminate the heating of our cryogenic system. To understand how the zinc source works in the presence of the helium film at low UV power, we have investigated whether electron trapping occurs.

\section{Electron source construction and measurement setup}
\label{sec:setup}

The photoemission source consisted of a low power pulsed xenon arc lamp~\cite{OceanOpticsLamp}, a \SI{600}{\micro\meter} core solarization resistant optical fiber~\cite{OceanOpticsFiber}, and a \SI{0.5}{\mm} thick piece of sapphire coated with a film of zinc. Since the transmittance of most optical fibers degrades when guiding UV light, a solarization resistant fiber was used because it transmits UV without degrading. The fiber was polished at both ends in order to achieve maximum transmittance. One end of the fiber was illuminated by the xenon lamp; the other end of the fiber was brought into the low temperature cell through a Stycast 2850 epoxy seal. In order for this seal to hold vacuum, the polyamide buffer coating on the fiber was stripped off using hot ($\sim \SI{100}{\celsius}$) sulfuric acid.\begin{figure}
\centering
  \includegraphics{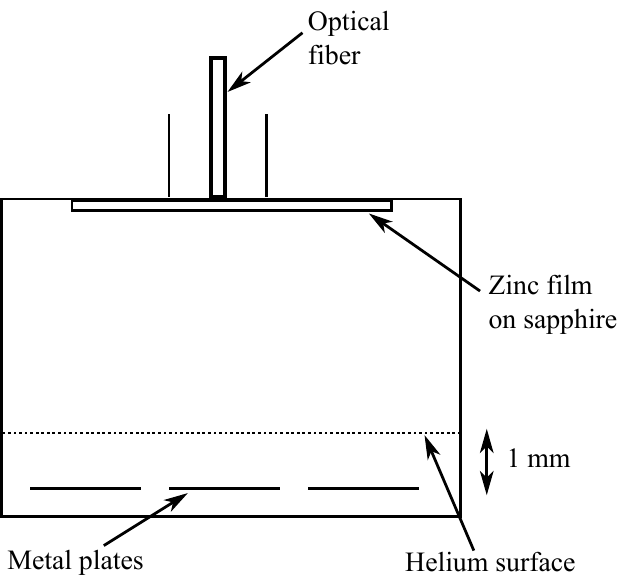}
\caption{Schematic of the photoemission source in low temperature cell along with the measurement plates for detecting electrons}
\label{fig:cell schematic}
\end{figure} As shown schematically in Fig.~\ref{fig:cell schematic}, the fiber illuminates a sapphire plate that has been coated with a zinc film on the bottom surface.

The film was made in a thermal evaporator by first depositing a \SI{10}{\nm} wetting layer of titanium followed by \SI{20}{\nm} of zinc. The resulting film was semi-transparent, while at the same time being conducting. The sapphire was held in place at the top of the cell by a Teflon holder and a copper clamp, which provided an electrical connection to the zinc independent of the cell walls. This independent electrical connection was useful to discharge electrons from the liquid helium surface by applying a positive voltage to the zinc. The zinc film was found to oxidize when exposed to air at room temperature, degrading the photoemission efficiency by about \SIrange[repeatunits=false,trapambigrange=false]{10}{20}{\percent} in an hour. To avoid oxidation, the zinc coated sapphire was vacuum sealed within the low temperature cell in less than about $\SI{30}{\minute}$ and then purged with helium gas.

The bottom half of the low temperature cell (Fig.~\ref{fig:cell schematic}) has 3 metal plates immersed \SI{1}{\mm} under the liquid helium surface. Electrons floating on the helium are detected by these plates using the Sommer--Tanner measurement scheme~\cite{Sommer1971}. A \SI{100}{\kHz} AC voltage is applied to the left plate. The center plate is at AC ground. The voltage on the right plate is measured by a lockin amplifier. While the helium surface is uncharged, any crosstalk voltage between plates was zeroed. When the helium surface is charged, a voltage appears on the right plate due to capacitive coupling between electrons on the helium surface and the metal plate. This voltage measured on the lockin is used to detect the presence of electrons.

\section{Electron source results}
\label{sec:results}

The zinc source was first tested under vacuum at room and liquid nitrogen temperature. The xenon lamp emits a \SI{3}{\micro\second} flash on each pulse carrying about \SI{100}{\nano\joule} of total energy. With the lamp pulsed at \SI{200}{\Hz}, the photoemitted current was measured to be \SI{100}{\pico\ampere} independent of temperature corresponding  to an electron emission rate of about \num{3e6} electrons per lamp pulse. The source was then tested at \SI{1.5}{\kelvin} with liquid helium in the cell, by charging the liquid helium surface with electrons and detecting them using the Sommer--Tanner scheme.\begin{figure}
\centering
  \includegraphics{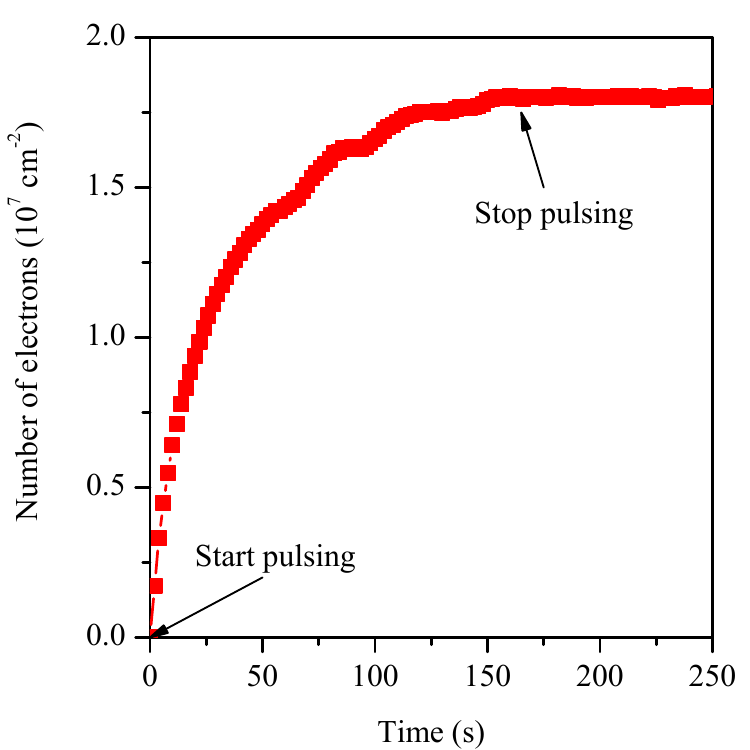}
\caption{Number of electrons on helium surface as a function of time showing charge up of the surface when the lamp is pulsed at \SI{200}{\Hz}. The initial slope of the charge up is about \SI{9e5}{\per\cms\per\second}. Line is a guide to the eye.}
\label{fig:charge up}
\end{figure} A charging graph like the typical one shown in Fig.~\ref{fig:charge up} indicates that electrons are being collected on the liquid helium surface.

The rate at which electrons were emitted by the source can be extracted from Fig.~\ref{fig:charge up}. When the lamp was turned on, the electron density on the helium surface first increased and then saturated since electrons already on the surface repel subsequent electrons that are emitted from the source. Upon saturation, there is no electric field between the zinc and the helium surface, and hence all the applied voltage is dropped across the helium above the metal plates. The electron density when the surface is saturated can be calculated from the capacitance of the helium above the metal plates~\cite{Stan1989} and is given by $n =\epsilon V/e d$ where $n$ is the electron density, $\epsilon$ is the permittivity of helium, $V$ is the applied DC voltage, $e$ is the electron charge and $d$ is the helium depth. The saturated electron density for our helium depths (\SI{1}{\mm}) and voltages (\SI{3}{\volt}) is calculated to be about \num{1.7e7} electrons/\si{\cms}. Assuming our signal arises from all electrons, we can calibrate the measured voltage in electrons/\si{\cms}. Consequently from Fig.~\ref{fig:charge up}, we see that when firing the lamp at \SI{200}{\Hz} over a surface area of about \SI{1}{\cms} the initial charging rate was about \SI{9e5} electrons/\si{\second} or a current of about \SI{0.1}{\pico\ampere}. Our source is therefore emitting electrons at a rate of about \num{5e3} electrons per \si{\cms} per lamp pulse in the presence of helium. Hence a desired number of electrons can be collected on the helium surface simply by controlling the number of lamp pulses.

\section{Photoemission mechanism}
\label{sec:mechanism}

Having shown that this photoemission source works, we then tried to understand the emission mechanism. The zinc source is covered by a thin film of superfluid helium, about \SI{35}{\nm} thick, that presents a barrier of about \SI{1}{\eV} to electrons being emitted~\cite{Woolf1965}. The mass of this film given by the volume (\SI{35}{\nm} by \SI{1}{\mm\squared}) times the density (\SI{0.145}{\gram\per\cmc}) is roughly \SI{5e-9}{\gram}. Assuming a latent heat of vaporization of \SI{90}{\joule\per\mole}, a rough estimate of the energy required to boil of this film is \SI{120}{\nano\joule}. Therefore unless we have the unlikely event that all the energy transmitted by the fiber (\SI{100}{\nano\joule}) gets dissipated in the helium film, we can safely assume that the film is present during the photoemission process. Further, the electron emission rate decreased in the presence of helium (\SI{0.1}{\pico\ampere}) when compared to vacuum (\SI{100}{\pico\ampere}) suggesting that the helium film definitely plays a role in how the source works.

There are three possible mechanisms for a photo-emitted electron to escape through the thin helium film and reach the bottom helium surface: electrons might be emitted with kinetic energy greater than the \SI{1}{\eV} barrier; electrons might tunnel through the thin \SI{35}{\nm} film; or, as suggested by Wilen and Gianetta~\cite{Wilen1985}, the electrons could be trapped and re-emitted from bubble states in the helium film. Electron trapping times in bubble states have been measured~\cite{Williams1971} to be at least about \SI{1}{\milli\second}. Therefore of these three possibilities, bubble formation can be detected by measuring how long after the optical pulse the electrons arrive at the helium surface at the bottom of the cell.

\begin{figure}
\centering
  \includegraphics{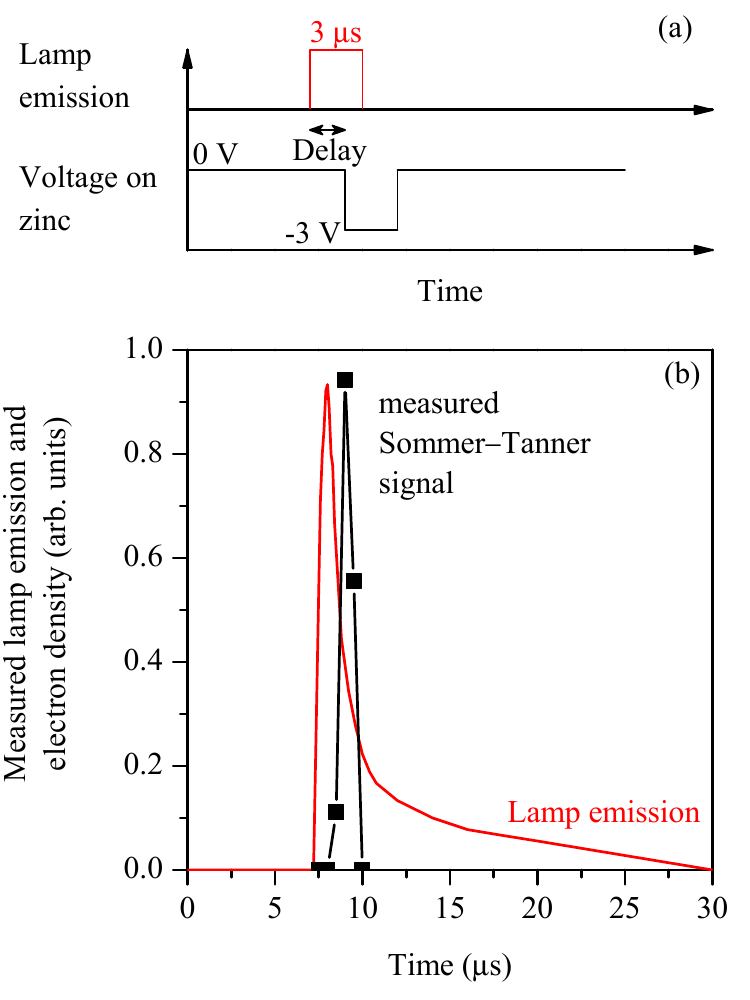}
\caption{(a) Timing diagram of \SI{3}{\micro\second} lamp emission and voltage applied to the zinc. (b) Experimentally determined lamp emission and measured Sommer--Tanner signal (squares) as a function of time. Lines are a guide to the eye. The initial, approximately \SI{1}{\micro\second}, time delay between the lamp emission and the arrival of electrons at the bottom helium surface, is due to scattering by the helium vapor in the cell. Lack of a \SI{1}{\milli\second} long tail indicates the absence of bubble formation}
\label{fig:time resolved}
\end{figure} A time-resolved experiment to detect the arrival of electrons on the bottom helium surface is sketched in Fig.~\ref{fig:time resolved}(a). A \SI{3}{\micro\second} long pulsed voltage waveform was applied to the zinc while holding the metal plates and cell walls at \SI{0}{\volt}. If the voltage on the zinc is \SI{0}{\volt} when the lamp emission occurs, no electrons are collected at the helium surface, whereas electrons are collected if the voltage is repulsive (\SI{-3}{\volt}). The lamp was pulsed \num{200} times, and the number of electrons collected on the helium was detected using the Sommer--Tanner scheme. The helium surface was then discharged before the next run by taking the zinc attractive. In each run, the delay time between the lamp pulse and the voltage pulse was varied, and therefore a time response of the electron source was mapped out, shown in Fig.~\ref{fig:time resolved}(b).

The shape of the time dependence of electrons arriving at the bottom surface looks similar to the shape of the optical pulse except for an $\sim \SI{1}{\micro\second}$ initial delay. The initial time delay in electron arrival after the optical pulse (Fig.~\ref{fig:time resolved}) is explained by noting that electrons can be scattered by helium gas atoms present within the cell at \SI{1.5}{\kelvin}. The mobility of electrons in low density helium gas has been measured previously~\cite{Levine1967} and found to follow the kinetic theory giving a mobility of \SI{1.8e5}{\cms\per\volt\per\second} at \SI{1.5}{\kelvin}. In the applied field of \SI{3}{\volt} over \SI{5}{\mm}, the scattering length of electrons is of the order of \SI{10}{\nm} and the time taken for electrons to drift from the zinc source to the helium surface is calculated to be of the order of \SI{1}{\micro\second}, matching the initial time delay in the time response of the electron source.

A long tail of the order of \SI{1}{\milli\second} in the time response would be indicative of trapping - probably by bubbles. However, as shown in Fig.~\ref{fig:time resolved}, no tail is detected in the electron time response. Except for an initial time delay, the shape of the electron pulse is very similar to that of the optical pulse suggesting that electrons are not trapped.

\section{Conclusion}
\label{sec:conclusion}
The photoemission source implemented in this paper was found to reliably produce electrons without heating our cryogenic system. About \num{5e3} electrons were collected per \si{\cms} of helium surface per lamp flash, and the total number of electrons collected could be controlled by varying the number of lamp flashes. Thus, our source fulfils all requirements for a good electron source to generate electrons on liquid helium. Additionally, we have not detected any electron trapping in the helium film covering the source indicating that the zinc source probably works by photoemission of electrons with kinetic energy above the \SI{1}{\eV} helium film barrier or by tunneling through the helium film. We find that this source is ideal for experiments involving a small number of electrons on helium at low temperatures such as those proposed for quantum computing~\cite{Sabouret2007}.

\begin{acknowledgments}
Supported by the National Science Foundation through the EMT program under Grant No. CCF-0726490
\end{acknowledgments}

\bibliography{photoem_references}

\begin{thebibliography}{10}%
\makeatletter
\providecommand \@ifxundefined [1]{%
 \ifx #1\undefined \expandafter \@firstoftwo
 \else \expandafter \@secondoftwo
\fi
}%
\providecommand \@ifnum [1]{%
 \ifnum #1\expandafter \@firstoftwo
 \else \expandafter \@secondoftwo
\fi
}%
\providecommand \enquote [1]{``#1''}%
\providecommand \bibnamefont  [1]{#1}%
\providecommand \bibfnamefont [1]{#1}%
\providecommand \citenamefont [1]{#1}%
\providecommand\href[0]{\@sanitize\@href}%
\providecommand\@href[1]{\endgroup\@@startlink{#1}\endgroup\@@href}%
\providecommand\@@href[1]{#1\@@endlink}%
\providecommand \@sanitize [0]{\begingroup\catcode`\&12\catcode`\#12\relax}%
\@ifxundefined \pdfoutput {\@firstoftwo}{%
 \@ifnum{\z@=\pdfoutput}{\@firstoftwo}{\@secondoftwo}%
}{%
 \providecommand\@@startlink[1]{\leavevmode\special{html:<a href="#1">}}%
 \providecommand\@@endlink[0]{\special{html:</a>}}%
}{%
 \providecommand\@@startlink[1]{%
  \leavevmode
  \pdfstartlink
   attr{/Border[0 0 1 ]/H/I/C[0 1 1]}%
   user{/Subtype/Link/A<</Type/Action/S/URI/URI(#1)>>}%
  \relax
 }%
 \providecommand\@@endlink[0]{\pdfendlink}%
}%
\providecommand \url  [0]{\begingroup\@sanitize \@url }%
\providecommand \@url [1]{\endgroup\@href {#1}{\urlprefix}}%
\providecommand \urlprefix [0]{URL }%
\providecommand \Eprint[0]{\href }%
\@ifxundefined \urlstyle {%
  \providecommand \doi [1]{doi:\discretionary{}{}{}#1}%
}{%
  \providecommand \doi [0]{doi:\discretionary{}{}{}\begingroup
  \urlstyle{rm}\Url }%
}%
\providecommand \doibase [0]{http://dx.doi.org/}%
\providecommand \Doi[1]{\href{\doibase#1}}%
\providecommand \bibAnnote [3]{%
  \BibitemShut{#1}%
  \begin{quotation}\noindent
    \textsc{Key:}\ #2\\\textsc{Annotation:}\ #3%
  \end{quotation}%
}%
\providecommand \bibAnnoteFile [2]{%
  \IfFileExists{#2}{\bibAnnote {#1} {#2} {\input{#2}}}{}%
}%
\providecommand \typeout [0]{\immediate \write \m@ne }%
\providecommand \selectlanguage [0]{\@gobble}%
\providecommand \bibinfo [0]{\@secondoftwo}%
\providecommand \bibfield [0]{\@secondoftwo}%
\providecommand \translation [1]{[#1]}%
\providecommand \BibitemOpen[0]{}%
\providecommand \bibitemStop [0]{}%
\providecommand \bibitemNoStop [0]{.\EOS\space}%
\providecommand \EOS [0]{\spacefactor3000\relax}%
\providecommand \BibitemShut [1]{\csname bibitem#1\endcsname}%
\bibitem{Ando1982}%
  \BibitemOpen
  \bibfield{author}{%
  \bibinfo {author} {\bibfnamefont{T.}~\bibnamefont{Ando}}, \bibinfo {author}
  {\bibfnamefont{A.~B.}\ \bibnamefont{Fowler}},\ and\ \bibinfo {author}
  {\bibfnamefont{F.}~\bibnamefont{Stern}},\ }%
  \bibfield{journal}{%
  \bibinfo {journal} {Rev. Mod. Phys.}\ }%
  \textbf{\bibinfo {volume} {54}},\ \bibinfo {pages} {437} (\bibinfo {year}
  {1982})%
  \bibAnnoteFile{NoStop}{Ando1982}%
\bibitem{Andrei1997}%
  \BibitemOpen
  \emph{\bibinfo {title} {{Two-dimensional electron systems on helium and other
  cryogenic substrates}}},\ edited by\ \bibinfo {editor} {\bibfnamefont{E.~Y.}\
  \bibnamefont{Andrei}}\ (\bibinfo {publisher} {Springer},\ \bibinfo {year}
  {1997})%
  \bibAnnoteFile{NoStop}{Andrei1997}%
\bibitem{Platzman1999}%
  \BibitemOpen
  \bibfield{author}{%
  \bibinfo {author} {\bibfnamefont{P.~M.}\ \bibnamefont{Platzman}}\ and\
  \bibinfo {author} {\bibfnamefont{M.~I.}\ \bibnamefont{Dykman}},\ }%
  \bibfield{journal}{%
  \bibinfo {journal} {Science}\ }%
  \textbf{\bibinfo {volume} {284}},\ \bibinfo {pages} {1967} (\bibinfo {year}
  {1999})%
  \bibAnnoteFile{NoStop}{Platzman1999}%
\bibitem{Dykman2003}%
  \BibitemOpen
  \bibfield{author}{%
  \bibinfo {author} {\bibfnamefont{M.~I.}\ \bibnamefont{Dykman}}, \bibinfo
  {author} {\bibfnamefont{P.~M.}\ \bibnamefont{Platzman}},\ and\ \bibinfo
  {author} {\bibfnamefont{P.}~\bibnamefont{Seddighrad}},\ }%
  \bibfield{journal}{%
  \bibinfo {journal} {Phys. Rev. B}\ }%
  \textbf{\bibinfo {volume} {67}},\ \bibinfo {pages} {155402} (\bibinfo {year}
  {2003})%
  \bibAnnoteFile{NoStop}{Dykman2003}%
\bibitem{Dahm2002}%
  \BibitemOpen
  \bibfield{author}{%
  \bibinfo {author} {\bibfnamefont{A.~J.}\ \bibnamefont{Dahm}}, \bibinfo
  {author} {\bibfnamefont{J.~M.}\ \bibnamefont{Goodkind}}, \bibinfo {author}
  {\bibfnamefont{I.}~\bibnamefont{Karakurt}},\ and\ \bibinfo {author}
  {\bibfnamefont{S.}~\bibnamefont{Pilla}},\ }%
  \bibfield{journal}{%
  \bibinfo {journal} {J. Low Temp. Phys.}\ }%
  \textbf{\bibinfo {volume} {126}},\ \bibinfo {pages} {709} (\bibinfo {year}
  {2002})%
  \bibAnnoteFile{NoStop}{Dahm2002}%
\bibitem{Papageorgiou2005}%
  \BibitemOpen
  \bibfield{author}{%
  \bibinfo {author} {\bibfnamefont{G.}~\bibnamefont{Papageorgiou}}, \bibinfo
  {author} {\bibfnamefont{P.}~\bibnamefont{Glasson}}, \bibinfo {author}
  {\bibfnamefont{K.}~\bibnamefont{Harrabi}}, \bibinfo {author}
  {\bibfnamefont{V.}~\bibnamefont{Antonov}}, \bibinfo {author}
  {\bibfnamefont{E.}~\bibnamefont{Collin}}, \bibinfo {author}
  {\bibfnamefont{P.}~\bibnamefont{Fozooni}}, \bibinfo {author}
  {\bibfnamefont{P.~G.}\ \bibnamefont{Frayne}}, \bibinfo {author}
  {\bibfnamefont{M.~J.}\ \bibnamefont{Lea}}, \bibinfo {author}
  {\bibfnamefont{D.~G.}\ \bibnamefont{Rees}},\ and\ \bibinfo {author}
  {\bibfnamefont{Y.}~\bibnamefont{Mukharsky}},\ }%
  \bibfield{journal}{%
  \bibinfo {journal} {Appl. Phys. Lett.}\ }%
  \textbf{\bibinfo {volume} {86}},\ \bibinfo {pages} {153106} (\bibinfo {year}
  {2005})%
  \bibAnnoteFile{NoStop}{Papageorgiou2005}%
\bibitem{Lyon2006}%
  \BibitemOpen
  \bibfield{author}{%
  \bibinfo {author} {\bibfnamefont{S.~A.}\ \bibnamefont{Lyon}},\ }%
  \bibfield{journal}{%
  \bibinfo {journal} {Phys. Rev. A}\ }%
  \textbf{\bibinfo {volume} {74}},\ \bibinfo {pages} {052338} (\bibinfo {year}
  {2006})%
  \bibAnnoteFile{NoStop}{Lyon2006}%
\bibitem{Sabouret2008}%
  \BibitemOpen
  \bibfield{author}{%
  \bibinfo {author} {\bibfnamefont{G.}~\bibnamefont{Sabouret}}, \bibinfo
  {author} {\bibfnamefont{F.~R.}\ \bibnamefont{Bradbury}}, \bibinfo {author}
  {\bibfnamefont{S.}~\bibnamefont{Shankar}}, \bibinfo {author}
  {\bibfnamefont{J.~A.}\ \bibnamefont{Bert}},\ and\ \bibinfo {author}
  {\bibfnamefont{S.~A.}\ \bibnamefont{Lyon}},\ }%
  \bibfield{journal}{%
  \bibinfo {journal} {Appl. Phys. Lett.}\ }%
  \textbf{\bibinfo {volume} {92}},\ \bibinfo {pages} {082104} (\bibinfo {year}
  {2008})%
  \bibAnnoteFile{NoStop}{Sabouret2008}%
\bibitem{Rees2010}%
  \BibitemOpen
  \bibfield{author}{%
  \bibinfo {author} {\bibfnamefont{D.}~\bibnamefont{Rees}}\ and\ \bibinfo
  {author} {\bibfnamefont{K.}~\bibnamefont{Kono}},\ }%
  \bibfield{journal}{%
  \bibinfo {journal} {J. Low Temp. Phys.}\ }%
  \textbf{\bibinfo {volume} {158}},\ \bibinfo {pages} {301} (\bibinfo {year}
  {2010})%
  \bibAnnoteFile{NoStop}{Rees2010}%
\bibitem{Paalanen1985}%
  \BibitemOpen
  \bibfield{author}{%
  \bibinfo {author} {\bibfnamefont{M.~A.}\ \bibnamefont{Paalanen}}\ and\
  \bibinfo {author} {\bibfnamefont{Y.}~\bibnamefont{Iye}},\ }%
  \bibfield{journal}{%
  \bibinfo {journal} {Phys. Rev. Lett.}\ }%
  \textbf{\bibinfo {volume} {55}},\ \bibinfo {pages} {1761} (\bibinfo {year}
  {1985})%
  \bibAnnoteFile{NoStop}{Paalanen1985}%
\bibitem{Williams1971}%
  \BibitemOpen
  \bibfield{author}{%
  \bibinfo {author} {\bibfnamefont{R.}~\bibnamefont{Williams}}, \bibinfo
  {author} {\bibfnamefont{R.~S.}\ \bibnamefont{Crandall}},\ and\ \bibinfo
  {author} {\bibfnamefont{A.~H.}\ \bibnamefont{Willis}},\ }%
  \bibfield{journal}{%
  \bibinfo {journal} {Phys. Rev. Lett.}\ }%
  \textbf{\bibinfo {volume} {26}},\ \bibinfo {pages} {7} (\bibinfo {year}
  {1971})%
  \bibAnnoteFile{NoStop}{Williams1971}%
\bibitem{Sommer1964}%
  \BibitemOpen
  \bibfield{author}{%
  \bibinfo {author} {\bibfnamefont{W.~T.}\ \bibnamefont{Sommer}},\ }%
  \bibfield{journal}{%
  \bibinfo {journal} {Phys. Rev. Lett.}\ }%
  \textbf{\bibinfo {volume} {12}},\ \bibinfo {pages} {271} (\bibinfo {year}
  {1964})%
  \bibAnnoteFile{NoStop}{Sommer1964}%
\bibitem{Sommer1971}%
  \BibitemOpen
  \bibfield{author}{%
  \bibinfo {author} {\bibfnamefont{W.~T.}\ \bibnamefont{Sommer}}\ and\ \bibinfo
  {author} {\bibfnamefont{D.~J.}\ \bibnamefont{Tanner}},\ }%
  \bibfield{journal}{%
  \bibinfo {journal} {Phys. Rev. Lett.}\ }%
  \textbf{\bibinfo {volume} {27}},\ \bibinfo {pages} {1345} (\bibinfo {year}
  {1971})%
  \bibAnnoteFile{NoStop}{Sommer1971}%
\bibitem{Klier2000}%
  \BibitemOpen
  \bibfield{author}{%
  \bibinfo {author} {\bibfnamefont{J.}~\bibnamefont{Klier}}, \bibinfo {author}
  {\bibfnamefont{I.}~\bibnamefont{Doicescu}},\ and\ \bibinfo {author}
  {\bibfnamefont{P.}~\bibnamefont{Leiderer}},\ }%
  \bibfield{journal}{%
  \bibinfo {journal} {J. Low Temp. Phys.}\ }%
  \textbf{\bibinfo {volume} {121}},\ \bibinfo {pages} {603} (\bibinfo {year}
  {2000})%
  \bibAnnoteFile{NoStop}{Klier2000}%
\bibitem{Wilen1985}%
  \BibitemOpen
  \bibfield{author}{%
  \bibinfo {author} {\bibfnamefont{L.~A.}\ \bibnamefont{Wilen}}\ and\ \bibinfo
  {author} {\bibfnamefont{R.~W.}\ \bibnamefont{Giannetta}},\ }%
  \bibfield{journal}{%
  \bibinfo {journal} {Rev. Sci. Instr.}\ }%
  \textbf{\bibinfo {volume} {56}},\ \bibinfo {pages} {2175} (\bibinfo {year}
  {1985})%
  \bibAnnoteFile{NoStop}{Wilen1985}%
\bibitem{OceanOpticsLamp}%
  \BibitemOpen
  \bibfield{author}{%
  \bibinfo {author} {\bibnamefont{{Ocean Optics PX-2 pulsed xenon lamp from
  Ocean Optics, 830 Douglas Ave, Dunedin, FL 34698.}}},\ }%
  \url{http://www.oceanoptics.com/Products/px2.asp}%
  \bibAnnoteFile{NoStop}{OceanOpticsLamp}%
\bibitem{OceanOpticsFiber}%
  \BibitemOpen
  \bibfield{author}{%
  \bibinfo {author} {\bibnamefont{{600 micron core solarization resistant fiber
  (FIBER-600-SR) from Ocean Optics, 830 Douglas Ave, Dunedin, FL 34698.}}},\ }%
  \url{http://www.oceanoptics.com/Products/fiberunjacketedbulk.asp}%
  \bibAnnoteFile{NoStop}{OceanOpticsFiber}%
\bibitem{Stan1989}%
  \BibitemOpen
  \bibfield{author}{%
  \bibinfo {author} {\bibfnamefont{M.~A.}\ \bibnamefont{Stan}}\ and\ \bibinfo
  {author} {\bibfnamefont{A.~J.}\ \bibnamefont{Dahm}},\ }%
  \bibfield{journal}{%
  \bibinfo {journal} {Phys. Rev. B}\ }%
  \textbf{\bibinfo {volume} {40}},\ \bibinfo {pages} {8995} (\bibinfo {year}
  {1989})%
  \bibAnnoteFile{NoStop}{Stan1989}%
\bibitem{Woolf1965}%
  \BibitemOpen
  \bibfield{author}{%
  \bibinfo {author} {\bibfnamefont{M.~A.}\ \bibnamefont{Woolf}}\ and\ \bibinfo
  {author} {\bibfnamefont{G.~W.}\ \bibnamefont{Rayfield}},\ }%
  \bibfield{journal}{%
  \bibinfo {journal} {Phys. Rev. Lett.}\ }%
  \textbf{\bibinfo {volume} {15}},\ \bibinfo {pages} {235} (\bibinfo {year}
  {1965})%
  \bibAnnoteFile{NoStop}{Woolf1965}%
\bibitem{Levine1967}%
  \BibitemOpen
  \bibfield{author}{%
  \bibinfo {author} {\bibfnamefont{J.~L.}\ \bibnamefont{Levine}}\ and\ \bibinfo
  {author} {\bibfnamefont{T.~M.}\ \bibnamefont{Sanders}},\ }%
  \bibfield{journal}{%
  \bibinfo {journal} {Phys. Rev.}\ }%
  \textbf{\bibinfo {volume} {154}},\ \bibinfo {pages} {138} (\bibinfo {year}
  {1967})%
  \bibAnnoteFile{NoStop}{Levine1967}%
\bibitem{Sabouret2007}%
  \BibitemOpen
  \bibfield{author}{%
  \bibinfo {author} {\bibfnamefont{G.}~\bibnamefont{Sabouret}},\ }%
  \emph{\bibinfo {title} {Towards spin-based quantum computing on liquid
  helium}},\ Ph.D. thesis,\ \bibinfo {school} {Princeton University} (\bibinfo
  {year} {2007})%
  \bibAnnoteFile{NoStop}{Sabouret2007}%
\end{thebibliography}%

\end{document}